\documentclass[aps,prc,twocolumn,superscriptaddress,showpacs,floatfix,nofootinbib]{revtex4-1}
\usepackage{amsmath,graphicx}
\def\p{{\bf p}}
\def\llangle{\left\langle}
\def\rrangle{\right\rangle}
\def\st{\begin{equation}}
\def\stp{\end{equation}}
\begin{document}

\title{Principal component analysis of event-by-event fluctuations}

\author{Rajeev S. Bhalerao}
\affiliation{Department of Theoretical Physics, Tata Institute of Fundamental Research,
Homi Bhabha Road, Mumbai 400005, India}
\author{Jean-Yves Ollitrault}
\affiliation{
CNRS, URA2306, IPhT, Institut de physique theorique de Saclay, F-91191
Gif-sur-Yvette, France} 
\author{Subrata Pal}
\affiliation{Department of Nuclear and Atomic Physics, Tata Institute of Fundamental Research, 
Homi Bhabha Road, Mumbai 400005, India}
\author{Derek Teaney}
\affiliation{Department of Physics and Astronomy, State University of New York, Stony Brook, NY 11794, USA}
\date{\today}

\begin{abstract}
We apply principal component analysis to the study of event-by-event fluctuations in 
relativistic heavy-ion collisions. 
This method brings out all the information contained in two-particle correlations in a physically transparent way. 
We present a guide to the method, 
and apply it to multiplicity fluctuations and anisotropic flow, using ALICE data and simulated events. 
In particular, we study elliptic and triangular flow fluctuations as a function of transverse momentum and rapidity. 
This method reveals previously unknown subleading modes in both
rapidity and transverse momentum for the momentum distribution as well
as elliptic and triangular flows.
\end{abstract}
\maketitle

\section{Introduction}
Anisotropic flow, $v_n$, is one of the most striking observations in 
nucleus-nucleus collisions at ultrarelativistic energies~\cite{Heinz:2013th,Gale:2013da}. 
It is an azimuthal asymmetry of particle production, which is interpreted as a signature of the 
system's hydrodynamic response to the initial density profile
of the overlap zone of the colliding nuclei.
The anisotropic flow thus provides a handle on the important issue of
thermalization of the quark-gluon matter formed in these collisions.
Event-by-event fluctuations of the initial density profile have long been recognized to play a 
crucial role in the interpretation of elliptic flow $v_2$~\cite{Alver:2006wh} and they are 
solely responsible for triangular flow $v_3$~\cite{Alver:2010gr}. 
However, the methods used to analyze anisotropic flow, namely the event-plane 
method~\cite{Poskanzer:1998yz} and cumulants~\cite{Borghini:2001vi}, 
were devised before the importance of fluctuations was recognized.
There are several discussions in the literature of how the existing
methods are affected by flow fluctuations~\cite{Adler:2002pu,Adams:2004bi,Alver:2008zza,Ollitrault:2009ie,Gombeaud:2009ye,Luzum:2012da}. 

We present a new method which unlike previous methods extracts the
flow fluctuations directly from data by fully exploiting all the
information contained in two-particle
correlations~\cite{Aamodt:2011by}. It also reveals an event-by-event
substructure in flow fluctuations whose various components are
organized by size thereby isolating the most important fluctuations.
These can be systematically analyzed by model calculations thus
providing additional constraints on the initial-state dynamics.

The flow picture is that particles are emitted independently with an underlying probability 
distribution which varies event to event~\cite{Luzum:2011mm}. 
In each event we write the single-particle distribution 
with $d\p \equiv dp_t \,d\eta\, d\varphi$
as 
\begin{equation}
\label{single}
\frac{dN}{d{\bf p}}=\sum_{n=-\infty}^{+\infty} V_n(p)e^{in\varphi} ,
\end{equation}
where 
$\varphi$ is the azimuthal angle of the outgoing particle momentum, 
$V_n(p)$ is a complex Fourier 
flow coefficient whose magnitude and phase fluctuate event to event, 
and $p$ is a shorthand notation for 
the remaining momentum coordinates, $p_{t}$ and $\eta$.  
$V_0(p)$ is real and corresponds to the momentum distribution, and $V_n^*=V_{-n}$.
Note that the usual definition of anisotropic flow $v_n$ is real and normalized: 
$v_n=|V_n|/V_0$. 

The covariance matrix of the flow harmonics $V_n(p)$ can be measured from the distribution of particle pairs.
Specifically,  in the flow picture
the pair distribution   is determined (predominantly) by the statistics  of the event-by-event single particle distribution 
\st
\left\langle\frac{dN_{\rm pairs}}{d\p_1 d\p_2}\right\rangle=
\llangle\frac{dN}{d\p_1 } \frac{dN}{d\p_2} \rrangle + \mathcal O (N),
\stp
where angular brackets denote an average over events, and 
the term $\mathcal O (N)$ corresponds to correlations not due to flow (``nonflow''), 
which are small for large systems. 

If the pair distribution 
is also expanded in a Fourier series
\begin{equation}
\label{pair}
\left\langle\frac{dN_{\rm pairs}}{d\p_1 d\p_2}\right\rangle=
\sum_{n=-\infty}^{+\infty} V_{n\Delta}(p_1,p_2) e^{in(\varphi_1 - \varphi_2) } \, , 
\end{equation}
then the measured series coefficients $V_{n\Delta}$ are determined by the statistics of $V_{n}$:
\st
\label{covariance}
V_{n\Delta}(p_1, p_2) =  \llangle V_{n}(p_1) V_n^*(p_2)  \rrangle   ,
\stp
where we have neglected nonflow correlations.\footnote{There is no
  systematic way of disentangling flow fluctuations and nonflow, unless specific
  assumptions are made~\cite{Luzum:2010sp,Xu:2012ue}.} 
The right-hand side of Eq.~(\ref{covariance}) is a covariance matrix, hence it is positive semidefinite. 
Thus a nontrivial property of flow correlations is that the measured pair correlation matrix $V_{n\Delta}(p_1, p_2)$ 
has only non-negative eigenvalues.\footnote{Back-to-back jets, on the other hand, typically result in 
large negative diagonal elements for odd $n$~\cite{Ollitrault:2012cm}, hence negative eigenvalues.}

The current Letter  uses the eigenmodes and eigenvalues of the 
two-particle correlation matrix, $V_{n\Delta}(p_1,p_2)$, 
to fully classify flow fluctuations in heavy-ion collisions.
Specifically, we show how a Principal Component Analysis (PCA)~\cite{PCAbook}
of $V_{n\Delta}(p_1,p_2)$ can be used to fully extract information on the pseudorapidity
and transverse-momentum-dependence
of flow fluctuations.  
We first test the applicability of the
method with Monte-Carlo simulations using the transport model 
AMPT~\cite{Lin:2004en} in both rapidity and transverse-momentum. 
In addition to the leading eigenmode, 
corresponding to the usual anisotropic flow (for $n>0$), 
the  correlation analysis reveals at least one important subleading
mode in both rapidity and $p_t$ for the momentum distribution and its
second and third harmonics.
Then we analyze ALICE data~\cite{Aamodt:2011by} in transverse-momentum 
and determine the first subleading elliptic
and triangular flow coefficients.

\section{Method}
Divide the detector acceptance into $N_b$ bins in transverse 
momentum and/or pseudorapidity, $p=(p_t,\eta)$.  The sample estimate for $V_n(p)$ in a 
given event (usually referred to as the flow vector~\cite{Poskanzer:1998yz})  is 
\begin{equation}
   Q_{n}(p) \equiv \frac{1}{2\pi \Delta p_t \Delta \eta }\sum^{M(p)}_{j=1} \exp(i n \varphi_j),
\end{equation}
where $M(p)$ 
is the number of particles in the bin and $\varphi_j$ is
the azimuthal angle of a particle. 
The pair distribution is 
\begin{multline}
\label{defvndelta}
V_{n\Delta}(p_1,p_2)
\equiv  \llangle  Q_{n}(p_1)  Q_{n}^*(p_2) \rrangle  - 
   \frac{\llangle M(p_1) \rrangle \delta_{p_1,p_2}}{(2\pi \Delta p_t \Delta \eta)^2 }  \\
  -\llangle  Q_{n}(p_1)\rrangle \llangle Q_{n}^*(p_2)\rrangle ,
\end{multline}
where the second term of the right-hand side subtracts 
self-correlations~\cite{Danielewicz:1985hn}. If self-correlations are not subtracted, $V_{n\Delta}(p_1,p_2)$ 
is positive semidefinite by construction. After subtraction, eigenvalues may have both signs. However, the eigenvalues will be positive 
if the correlations are due to collective flow. 

The last term on the right-hand side of Eq.~(\ref{defvndelta}) subtracts the mean value in order to single out the 
fluctuations. For $n>0$ and an azimuthally symmetric detector, this
term vanishes by azimuthal symmetry. 
For an asymmetric detector, subtracting the mean value corrects for azimuthal anisotropies in the acceptance~\cite{Borghini:2000sa,Bilandzic:2013kga}
Note that we define $V_{n\Delta}(p_1,p_2)$ as a {\it sum\/} over all pairs, as opposed to the usual 
normalization~\cite{Alver:2010gr,Aamodt:2011by} where one {\it averages\/} over pairs in each 
bin.\footnote{
The present normalization is required by the principal component analysis for $n=0$, and is desirable for $n=2,3$ because it gives weight to a bin that is of the order of the number of particles in it.
Note that a similar normalization is now used in cumulant analyses of anisotropic 
flow, where it is better to give all pairs (or generally multiplets) 
the same weight~\cite{Bhalerao:2003xf,Bilandzic:2010jr,Bilandzic:2013kga}.}

The principal component analysis approximates the pair distribution 
as:
\begin{equation}
\label{PCA}
   V_{n\Delta}(p_1,p_2) \approx \sum_{\alpha = 1}^k
   V_n^{(\alpha)}(p_1) V_n^{(\alpha)*}(p_2),
\end{equation}
where each term 
in the sum corresponds to a different component (mode) of flow fluctuations, and 
$k\le N_b$. 
If there are no flow fluctuations,  the pair distribution 
$V_{n\Delta}(p_1,p_2)$ factorizes~\cite{Aamodt:2011by} and 
there is only one component, \emph{i.e.} $k=1$ in Eq.~(\ref{PCA}),
corresponding to the usual anisotropic flow. 
Flow fluctuations break factorization~\cite{Gardim:2012im}. 
Higher-order principal components then reveal information about the 
statistics and momentum dependence of flow fluctuations. 

In practice, the principal components are obtained by diagonalizing
$V_{n\Delta}(p_1,p_2)=\sum_\alpha \lambda^{(\alpha)}\psi^{(\alpha)}(p_1)\psi^{(\alpha)*}(p_2)$  
(where $\psi^{(\alpha)}(p)$ denotes the normalized eigenvector) 
and ordering eigenvalues $\lambda^{(\alpha)}$ from largest to smallest, 
$\lambda^{(1)}>\lambda^{(2)}>\lambda^{(3)}\cdots$.
Identifying with Eq.~(\ref{PCA}), one obtains
\begin{equation}
\label{defPC}
V_n^{(\alpha)}(p)\equiv\sqrt{\lambda^{(\alpha)}}\psi^{(\alpha)}(p).
\end{equation}
Because of the square root, eigenvalues must be positive. 
If parity is conserved, the correlation matrix $V_{n\Delta}(p_1,p_2)$
is real up to statistical fluctuations, and $V_n^{(\alpha)}(p)$ can be
chosen to be real.

The flow in a given event can be written as
\begin{equation}
V_n(p)= \sum_{\alpha = 1}^k \xi^{(\alpha)}V_n^{(\alpha)}(p), 
\end{equation}
where $\xi^{(\alpha)}$ are complex, uncorrelated random variables with zero mean and unit variance, 
that is, $\langle \xi^{(\alpha)}\rangle=0$ and $\langle \xi^{(\alpha)}\xi^{(\beta)*}\rangle=\delta_{\alpha,\beta}$. 
The rms magnitude and momentum dependence of  flow fluctuations are determined
by the corresponding properties of the principal components. 
Since eigenmodes are real, the azimuthal angle of anisotropic flow 
in a specific event is solely determined by the phases of $\xi^{(\alpha)}$.

For sake of compatibility with the usual definition of $v_n(p)$ which is the anisotropy per particle, we define
\begin{equation}
v_n^{(\alpha)}(p)\equiv \frac{V_n^{(\alpha)}(p)}{\langle V_0(p)\rangle}. 
\end{equation}
Thus, $v_0^{(\alpha)}(p)$ describe relative multiplicity fluctuations, while $v_n^{(\alpha)}(p)$ describe fluctuations of anisotropic flow.

\section{Results}

In order to illustrate the method, we analyze $10^4$ Pb-Pb collisions at $\sqrt{s}=2.76$~TeV
in the 0-10\% centrality range,\footnote{We only show one centrality bin for sake of illustration, 
but we have checked that results are similar for other centralities.}  
generated using the string-melting version of the AMPT model~\cite{Lin:2004en}. 
Initial conditions are generated via the HIJING 2.0 model~\cite{Deng:2010mv}
which contains nontrivial event-by-event fluctuations at the nucleonic and 
partonic levels \cite{Pang:2012he}. 
In AMPT, collective flow is generated mainly as a result of partonic cascade.
AMPT also has resonance formations and decays,
and thus contains non-flow effects. 
We have checked that the present implementation reproduces LHC
data for anisotropic flow ($v_2$ to $v_6$) at all centralities~\cite{Pal:2012gf}.

We first construct the pair distribution, Eq.~(\ref{defvndelta}), for all particles in the $-3<\eta<3$ pseudorapidity window, 
in $\eta$ bins of $0.5$. 
We then diagonalize the $12\times12$ matrix corresponding to these pseudorapidity bins. 
The eigenvalues are in general strongly ordered from largest to smallest. There are a few negative 
eigenvalues which can be attributed to statistical 
fluctuations.\footnote{This can be checked by applying the 
PCA to purely statistical fluctuations. We generated random matrices 
according to the statistical error of $V_{n\Delta}(p_1,p_2)$,
and found that the negative eigenvalues of $V_{n\Delta}(p_1,p_2)$ 
are compatible with 
the  negative eigenvalues of these random matrices.}

\begin{figure}[ht]
\includegraphics[width=\linewidth]{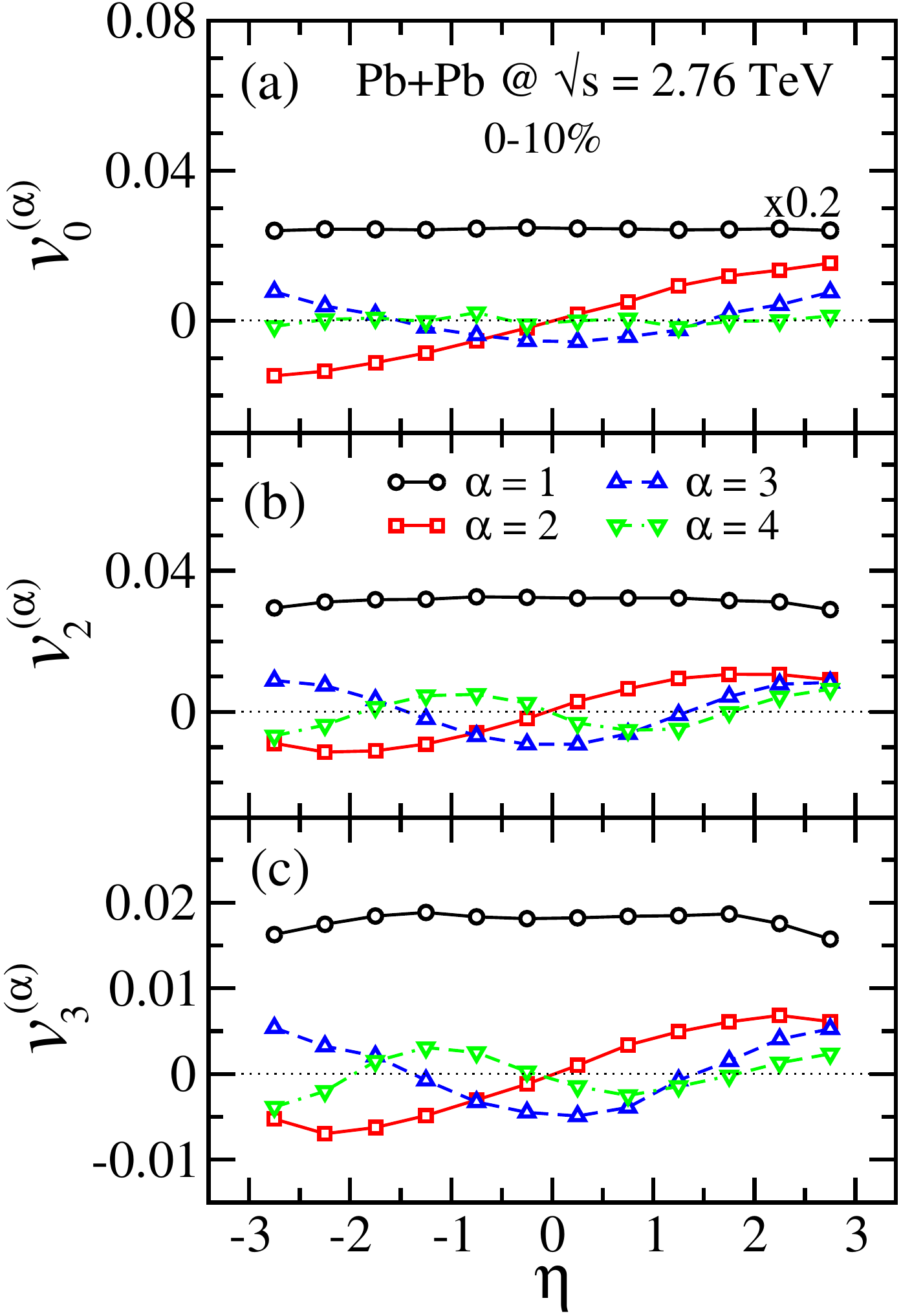}
\caption{Principal component analysis as a function of pseudorapidity for 
Pb+Pb collisions at $\sqrt{s}=2.76$~TeV in the 0-10\% centrality window 
generated with AMPT. 
(a) Multiplicity fluctuations; 
(b) Elliptic flow fluctuations; 
(c) Triangular flow fluctuations. 
}
\label{eta}
\end{figure}

The leading principal components for $n=0$, $n=2$, and $n=3$ are shown in Fig.~\ref{eta}. 
Fig.~\ref{eta} (a) displays the principal modes of multiplicity fluctuations ($n=0$) as a function of pseudorapidity. 
The leading mode $v_0^{(1)}(\eta)$ 
is a global $12\%$ relative fluctuation, independent of $\eta$, corresponding to the fluctuation of the total 
multiplicity within the event sample. 
The next-to-leading mode $v_0^{(2)}(\eta)$ is odd and of much smaller amplitude, as shown by the eigenvalues, $\lambda^{(2)}\sim \lambda^{(1)}/60$. 
A natural source of this odd mode is the small difference between the participant numbers of projectile and target
nuclei induced by fluctuations, which creates a forward-backward asymmetry of the 
multiplicity~\cite{Bzdak:2012tp,Vovchenko:2013viu}. 
Since both the colliding system and the analysis window are symmetric around $\eta=0$, 
principal components have definite parity in $\eta$, up to statistical fluctuations. Indeed, the next mode 
$v_0^{(3)}(\eta)$ is even, suggesting that principal components typically have alternating parities. 
The corresponding eigenvalue is again much smaller, $\lambda^{(3)}\sim \lambda^{(2)}/5$. 
$v_0^{(4)}(\eta)$ and higher modes are blurred by statistical fluctuations. 
Note that Eq.~(\ref{PCA}) defines principal components up to a sign. 
Here, we conventionally choose $v_n^{(\alpha)}(\eta)>0$ at forward rapidity.
Fig.~\ref{eta}  also illustrates the orthogonality of principal components, that is, 
\begin{equation}
\label{orthogonal}
\sum_\eta V_n^{(\alpha)}(\eta) V_n^{(\beta)*}(\eta)=0 \ \ {\rm if}\ \ \alpha\not=\beta.
\end{equation}
Thus, $v_n^{(\alpha)}(\eta)$ typically has $\alpha-1$ nodes.

Fig.~\ref{eta} (b) and (c) display the principal components of elliptic and triangular flow fluctuations as a function of pseudorapidity. 
The leading modes $v_n^{(1)}(\eta)$ correspond to the usual elliptic and triangular flows, which 
depend weakly on $\eta$ at the LHC~\cite{ATLAS:2011ah,Chatrchyan:2013kba}. 
The subleading modes $v_n^{(2)}(\eta)$ are odd and of smaller
amplitude ($\lambda^{(2)}\simeq \lambda^{(1)}/13$). These rapidity-odd
harmonic flows, or torqued flows, can be attributed to  
the small relative angle between $n$-th harmonic participant planes defined in the  projectile and target nuclei~\cite{Bozek:2010vz}. 

Note that the correlation matrix  $V_{n\Delta}(\eta_1,\eta_2)$ is the sum of
flow and nonflow correlations~\cite{Alver:2010rt}. 
The nonflow correlation is significant only for small values of the
relative pseudorapidity $\Delta\eta\equiv|\eta_1-\eta_2|$. If the
range in $\Delta\eta$ is smaller than the binning, it contributes to 
the diagonal elements, and its effect is to shift all eigenvalues by 
a constant. We observe in general a clear ordering of eigenvalues 
($\lambda^{(2)}/\lambda^{(3)}\sim \lambda^{(3)}/\lambda^{(4)}\sim 2$)
which suggests that the correlation has a long range in $\Delta \eta$ and is
therefore dominated by flow.   Visual inspection of correlation matrix $V_{n\Delta}(\eta_1,\eta_2)$ qualitatively confirms this reasoning.

We then carry out the analysis as a function of transverse momentum. 
In addition to AMPT generated events, 
we use experimental data for $V_{n\Delta}(p_1,p_2)$ provided by the ALICE collaboration~\cite{Aamodt:2011by}
for Pb+Pb collisions in the 0-10\% centrality window. 
ALICE uses all charged particles in the pseudorapidity window $|\eta|<1$. 
The definition of $V_{n\Delta}$ is not quite the same as ours: First, it is averaged (as opposed to summed) 
over pairs. We correct for this difference by multiplying 
off-diagonal elements of $V_{n\Delta}$ of ALICE
by the average multiplicity of pairs in each $(p_1,p_2)$ bin, 
which we estimate using the statistical errors provided by ALICE ($\sigma\simeq (2N_{\rm pairs})^{-1/2}$). The diagonal elements are multiplied
by twice the average multiplicity of the pairs, to account for 
self-correlations.
Second, the analysis is done with a rapidity gap to suppress nonflow
correlations: this means that particles 1 and 2 in Eq.~(\ref{defvndelta}) are 
separated by a rapidity gap of $0.8$. 
For sake of comparison, we repeat the analysis using AMPT events (the same as in Fig.~\ref{eta}). 
In order to compensate for the lower statistics, we use a wider pseudorapidity window, from 
$-2$ to $2$, with a $0.8$ rapidity gap between $-0.4$ and $0.4$. 
\begin{figure}[ht]
\includegraphics[width=\linewidth]{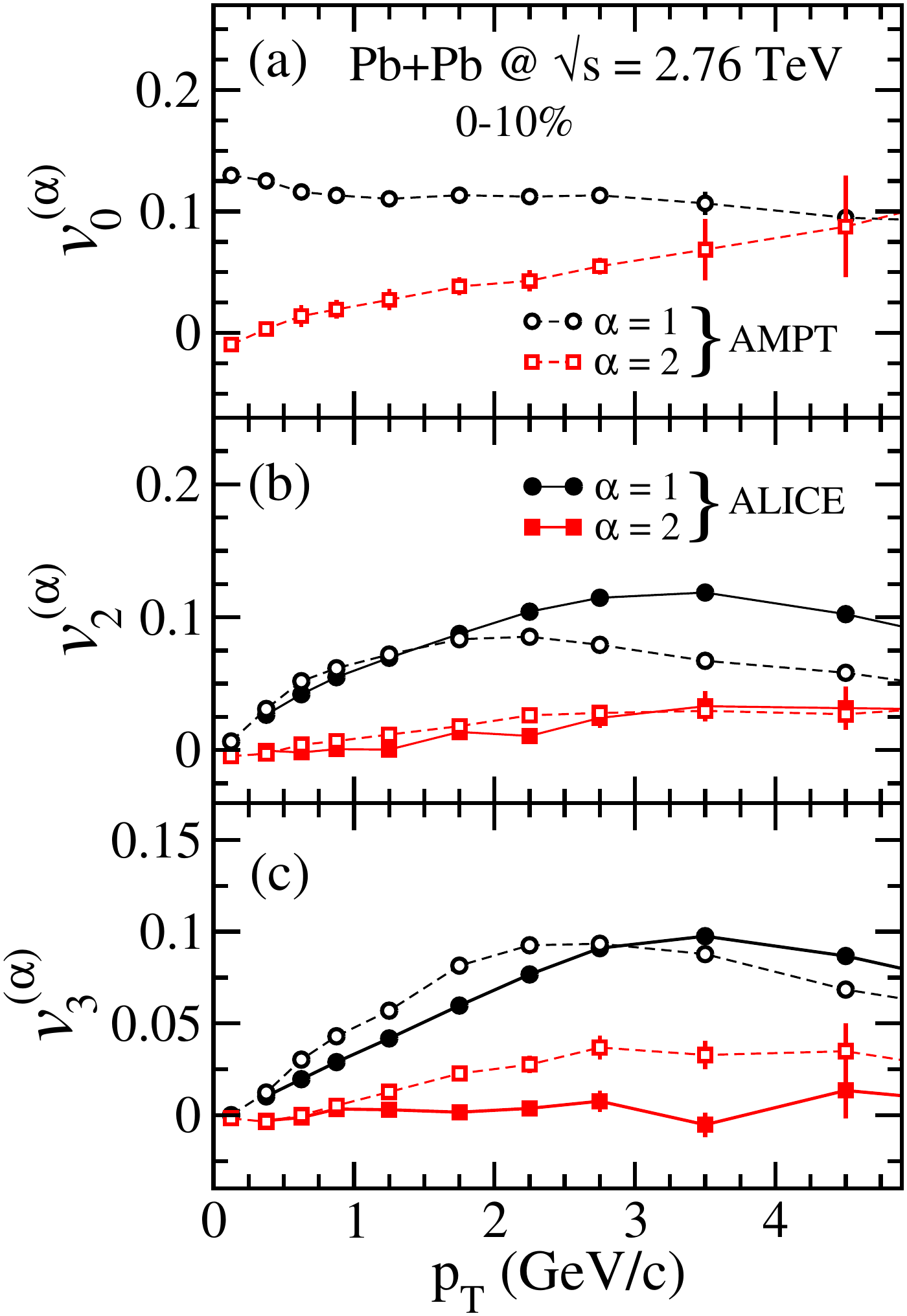}
  \caption{Principal component analysis as a function of transverse momentum for 
Pb+Pb collisions at $\sqrt{s}=2.76$~TeV in the 0-10\% centrality window, 
using ALICE data~\cite{Aamodt:2011by} (full symbols) and AMPT events (open symbols). 
(a) Multiplicity fluctuations; 
(b) Elliptic flow fluctuations; 
(c) Triangular flow fluctuations. 
 }
\label{pt}\end{figure}

Fig.~\ref{pt} (a) displays the two principal modes of multiplicity fluctuations ($n=0$) as a function of transverse momentum. Since no experimental data are available for this analysis, we only use AMPT events. 
As in Fig.~\ref{eta} (a), the leading mode is essentially constant and
corresponds to the 12\% fluctuation in the total multiplicity. The subleading
mode  increases linearly as a function of $p_t$ for large $p_t$, which can be
interpreted as the result of a radial flow fluctuation. Specifically, in a
hydrodynamic model the number of particles at high $p_t$ decreases
as, $\exp\left(p_t (u-u_0)/T\right)$,  where $u$ is the
maximum fluid 4-velocity, $u_0=\sqrt{1+u^2} $ and $T$ is the temperature~\cite{Borghini:2005kd}. A
small variation in $u$ therefore produces a relative variation in the yield
increasing linearly with $p_t$. 

Fig.~\ref{pt} (b) and (c) display the two leading principal components for elliptic and triangular flows. 
ALICE data show a very strong ordering between the two leading eigenvalues
($\lambda^{(1)}/\lambda^{(2)}\sim 400$ for $n=2$,  $300$ for $n=3$). 
The leading components for $n=2$ and $n=3$ are very close to the value of $v_2$ and $v_3$ obtained
using the same data~\cite{Aamodt:2011by}. 
The subleading mode is of much smaller magnitude and becomes significant 
only at large transverse momentum. 
In a hydrodynamical model, such a behavior is expected. 
Indeed, in a typical event the phase angle of the harmonic flow at high-momentum deviates slightly from the phase angle at low-momentum~\cite{Ollitrault:2012cm,Heinz:2013bua}.
Thus, the flow of high-$p_t$ particles has a small component independent 
of the flow of low-$p_t$ particles. The subleading mode determines the magnitude of this 
additional component. 
As we shall see below, it is also responsible for the factorization breaking of azimuthal 
correlations observed in these data~\cite{Aamodt:2011by,Gardim:2012im}.

\section{Discussion}

This new method, unlike traditional analysis methods, makes use of
{\it  all\/} the information contained in two-particle azimuthal
correlations. Specifically, it uses the detailed information on how
they depend on the momenta of {\it both\/} particles, 
as opposed to traditional analyses which integrate over one of the momenta. 

Previously, this double-differential structure has been used to test the factorization 
of azimuthal correlations. Small factorization breaking effects have been seen 
experimentally~\cite{Aamodt:2011by,CMS:2014vca} and in 
event-by-event hydrodynamic 
calculations~\cite{Gardim:2012im,Heinz:2013bua,Kozlov:2014fqa}. 
They have been characterized by the Pearson correlation coefficient between two different momenta: 
\begin{equation}
\label{rn}
r=\frac{V_{n\Delta}(p_1,p_2)}{\sqrt{V_{n\Delta}(p_1,p_1)V_{n\Delta}(p_2,p_2)}}. 
\end{equation}
$r=\pm 1$ if correlations factorize, and $|r|\le 1$ in general if correlations are due to flow.

These results are easily recovered in the language of principal components, in a physically 
transparent way.   
Factorization is the limiting case of just one principal component ($k=1$ in Eq.~(\ref{PCA})). 
In the more general case $k>1$, Eq.~(\ref{PCA}) guarantees $|r|\le 1$: 
Cauchy-Schwarz inequalities~\cite{Gardim:2012im} are equivalent to the condition that all eigenvalues of the matrix 
$V_{n\Delta}$ 
are positive. 
Flow fluctuations are typically dominated by a single subleading mode, i.e., $k=2$,
with $|V_n^{(2)}(p)|\ll |V_n^{(1)}(p)|$. In this limit, Eq.~(\ref{rn}) gives
\begin{equation}
1-r\simeq \frac{1}{2}\left|\frac{V_n^{(2)}(p_1)}{V_n^{(1)}(p_1)}- \frac{V_n^{(2)}(p_2)}{V_n^{(1)}(p_2)} \right|^2. 
\end{equation}
We see that $r\le 1$ as required by the Cauchy-Schwarz inequality. Further, 
we see that the breaking of factorization is induced by the relative difference between the subleading mode and the leading mode. 

To summarize, we have presented a new method to analyze the
anisotropic flow data in relativistic heavy-ion collisions. It is
based on the Principal Component Analysis applied to the two-particle
correlation matrix.
The principal components express the detailed information contained in correlations 
in a convenient way, which can be directly compared with  model calculations. 
The method has revealed for the first time subleading modes in both
rapidity and $p_t$ in the harmonic coefficients for $n=0,2$ and 3.
We anticipate a rich experimental and theoretical program  studying
the dynamics of these subleading flow vectors, which can be used to further
constrain the plasma  response to the initial geometry.

\begin{acknowledgments}
   We thank A. Bzdak for early discussions of this work.
This work is funded by CEFIPRA under project 4404-2.
JYO thanks the MIT LNS for hospitality,
and acknowledges support by the 
European Research Council under the Advanced Investigator Grant ERC-AD-267258. 
DT is supported by the U.S. Department of Energy grant DE-FG02-08ER4154.
\end{acknowledgments}

\end{document}